\documentclass[conference]{IEEEtran}

\usepackage[nolist]{acronym}
\usepackage{subcaption}
\usepackage{tikz}
\usetikzlibrary{shapes, arrows, calc, positioning}
\usepackage{pgfplots}
\usepackage{pgfplotstable}
\usepgfplotslibrary{fillbetween}
\pgfplotsset{compat=1.18}
\usepgfplotslibrary{groupplots}
\usepackage{tabularx}
\usepackage{cite}
\usepackage{url}
\usepackage{booktabs}
\usepackage[cmex10]{amsmath}
\usepackage[final]{changes}
\usepackage{hyperref}

\begin{document}

\title{On the Flexibility Potential of a Swiss Distribution
Grid: Opportunities and Limitations}

\author{
\IEEEauthorblockN{Jan Brändle\IEEEauthorrefmark{1}\IEEEauthorrefmark{2}, Julie Rousseau\IEEEauthorrefmark{1}\IEEEauthorrefmark{3}, Pulkit Nahata\IEEEauthorrefmark{1}, Gabriela Hug\IEEEauthorrefmark{1}}
\IEEEauthorblockA{\IEEEauthorrefmark{1}Power Systems Laboratory, ETH Zürich, Zürich, Switzerland}
\IEEEauthorblockA{\IEEEauthorrefmark{3} Urban Energy Systems Laboratory, Empa, D\"{u}bendorf, Switzerland}
\IEEEauthorblockA{\IEEEauthorrefmark{2}janbr@ethz.ch}
}

\begin{acronym}
  \acro{DSO}{Distribution System Operator}
  \acro{FOR}{Feasible Operating Region}
  \acro{FFOR}{Feasible Flexibility Operating Region}
  \acro{HV}{High Voltage}
  \acro{PCC}{Point of Common Coupling}
  \acro{PV}{Photovoltaic}
  \acro{OPF}{Optimal Power Flow}
  \acro{LP}{Linear Problem}
  \acro{DER}{Distributed Energy Resource}
  \acro{LV}{Low Voltage}
  \acro{MV}{Medium Voltage}
  \acro{BESS}{Battery Energy Storage System}
  \acro{SOC}{State of Charge}
  \acro{MPPT}{Maximum Power Point Tracking}
\end{acronym}

\maketitle

\pagestyle{plain}

\begin{abstract}
The growing integration of distributed renewable generation and the electrification of heating and transportation are rapidly increasing the number of flexible devices within modern distribution grids. Leveraging the aggregated flexibility of these small-scale distributed resources is essential to maintaining future grid-wide stability. This work uses the Swiss distribution grid of Walenstadt as a case study to provide insights into the aggregated flexibility potential of distribution grids. It demonstrates that incorporating devices such as heat pumps and photovoltaic systems significantly enhances distribution grid flexibility. It investigates the time-varying nature of aggregated flexibility and highlights how it can vary seasonally. Furthermore, simulations of future scenarios reveal that aggregated flexibility does not increase linearly or monotonically with higher levels of flexible device penetration. This is primarily due to the overloading of individual feeders, which underscores the impact of grid topology and network constraints on the aggregated flexibility potential.
\end{abstract}


\section{Introduction}
Electric power grids face significant challenges due to the rise in distributed generation and the electrification of heating and transportation. These trends lead to increased loads in the system and introduce greater variability and uncertainty in both power generation and consumption, which makes it more challenging to maintain system stability. Flexibility, defined as the ability of a device to deviate from its planned power set points, can help to maintain the balance between generation and consumption. Devices such as heat pumps, rooftop \ac{PV} installations, and battery systems have an inherent flexibility potential. Yet, their individual small scale often prevents this potential from being effectively harnessed. In order for system operators to utilize these resources, such as for grid support through participation in balancing markets, it is crucial to assess their aggregated flexibility potential at the distribution system level \cite{Koolen2023, rossi2020}. 

While determining flexibility is simple for individual devices, it becomes significantly more complex for entire networks. Determining aggregated distribution grid flexibility requires careful consideration of both device-specific limitations and network constraints, which include power flow limits and voltage limits, and are impacted by grid topology. A promising approach to characterize this overall flexibility is the computation of the \ac{FOR} of the distribution grid \added{\cite{silva2018a, contreras2018, gonzalez2018}}, which delimits feasible operating states in the active-reactive power (PQ) plane. 

Existing studies propose different methods for quantifying the \ac{FOR}. They can broadly be classified into Minkowski-sum-based methods \cite{ULBIG2015155, riaz2022}, Monte Carlo techniques \cite{riaz2019, contreras2019comparison}, or \ac{OPF}-based methodologies \cite{lopez2021quickflex, riaz2022, contreras2018}. Computing the Minkowski sum of individual \acp{FOR} results in an exact but computationally challenging aggregation. Besides, Minkowski-sum-based methods cannot, in general, integrate grid-level constraints, such as line constraints \cite{ULBIG2015155, riaz2022}. Monte Carlo techniques estimate the FOR by computing many power flow simulations for randomly selected operating points. They require running a large number of simulations, especially as the number of connected \acp{DER} increases \cite{riaz2019}, and may misrepresent and especially underestimate the FOR \cite{contreras2019comparison}. \ac{OPF}-based methodologies iteratively solve optimization problems to construct the \ac{FOR}. They use either nonlinear AC-OPF equations \added{\cite{riaz2022, lopez2021quickflex, capitanescu2018}}, which offer higher fidelity but pose computational and optimality challenges \cite{contreras2018}, or linearized power flow equations, which yield a simpler linear problem at the cost of reduced accuracy \cite{lopez2021quickflex, contreras2018}. In this work, we adopt the linearized OPF-based formulation for its ability to explicitly represent grid constraints and for its computational efficiency.

While these computational approaches are extensively researched, existing studies often rely on simulated and synthetic data or artificial network topologies \cite{riaz2019, lopez2021quickflex}, neglect network topology \cite{ULBIG2015155}, or do not consider time-varying characteristics of loads and generation \cite{riaz2022, contreras2019comparison}. This limits their ability to reflect the flexibility of real-world distribution grids.

To address these limitations and gain realistic insights into the flexibility potential of distribution grids, we collaborate with the Swiss utility of Walenstadt. Its grid is characterized by numerous \ac{PV} systems, high-capacity batteries, and flexible loads such as heat pumps, and therefore serves as an ideal test bed. \added{By applying the established flexibility quantification framework of the \ac{FOR} to this real-world case study, we bridge the gap between theoretical modeling and practical flexibility assessment. We provide researchers and grid operators with insights into the upper-bound of available distribution grid flexibility, and discuss its potential as well as its temporal variations and limitations. Such assessment supports grid management across multiple timeframes, ranging from real-time quantification of aggregated flexibility to long-term planning and evaluation of business cases for balancing market participation.} The contributions of this paper are threefold:

\begin{itemize}
    \item We design device-level models based on historical real-world consumption, generation, and device-level data to closely mimic the actual operation of a distribution grid.  
    \item  We compute and analyze the distribution grid flexibility of the Walenstadt \ac{MV} grid. In particular, we investigate the temporal variation of the grid's flexibility across days and seasons.
    \item We assess how the grid's flexibility may evolve in future scenarios in which more devices become electrified. We specifically highlight the role of topology and grid constraints in distribution grid flexibility.   
\end{itemize}
The remainder of this paper is structured as follows. Section~\ref{sec:methodology} first details the modeling of device-level and network-level constraints and then introduces the \ac{OPF}-based computation of the \ac{FFOR}, which is a reformulation of the \ac{FOR}. Section~\ref{sec:caseStudy} describes the \ac{MV} distribution grid operated by the utility of Walenstadt and details how we leverage historical data to infer missing information. Section~\ref{sec:results} demonstrates the application of our methodology in various scenarios and highlights key results. Finally, Section~\ref{sec:conclusion} summarizes the most important conclusions.

\section{Methodology}\label{sec:methodology}
In this paper, flexibility refers to a device's or a system's ability to deviate from its active and reactive power set points. For a distribution grid, these set points correspond to the planned active and reactive power exchange with the higher voltage grid at the \ac{PCC}. Consequently, the term distribution grid flexibility describes the collective ability of many individual flexible devices located in the distribution grid to deviate from their planned operation, resulting in a deviation from planned exchanges at the PCC. In this section, we introduce the modeling of the various components required to quantify distribution grid flexibility: the individual flexible devices, the distribution grid that connects them, and the OPF-based algorithm that computes the aggregated flexibility at the PCC.

\subsection{Individual Flexible Devices}

Some devices located in distribution grids can consume/produce power flexibly, e.g., \ac{BESS}, heat pumps, and PV systems. These devices can temporarily deviate from their baseline operation to adapt to the power system's needs. However, their flexibility can be subject to technical constraints imposed by the device itself and/or comfort constraints imposed by the users of the device. These constraints can restrict the device's active and/or reactive power output, or its energy consumption, i.e., the total amount of power that can be consumed over a time period. Here, we formalize the device-level constraints of three device types: \acp{BESS}, \ac{PV} systems, and controllable loads.

\begin{figure}[h]
     \centering
     \begin{subfigure}[b]{0.33\linewidth}
         \centering
         \includegraphics[height=2.88cm]{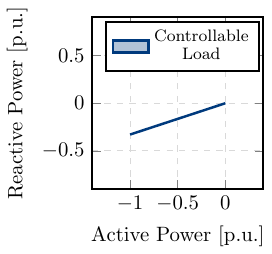}
         \caption{Controllable Load.}
         \label{fig:devices_cl}
     \end{subfigure}
     \hfill
     \begin{subfigure}[b]{0.3\linewidth}
         \centering
         \includegraphics[height=2.7cm]{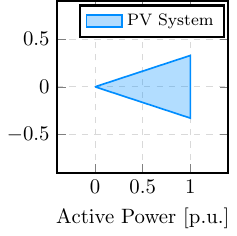}
         \caption{PV System.}
         \label{fig:devices_pv}
     \end{subfigure}
     \hfill
     \begin{subfigure}[b]{0.3\linewidth}
         \centering
         \includegraphics[height=2.7cm]{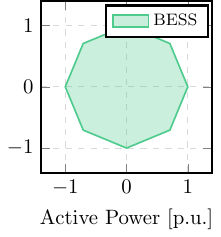}
         \caption{BESS.}
         \label{fig:devices_bess}
     \end{subfigure}
        \caption{\replaced{Capability charts}{FORs} of three individual flexible devices.}
        \label{fig:for_devices}
\end{figure}

\subsubsection{\ac{BESS}}
\label{subsec:device_bess}
For both residential and large-scale \ac{BESS}, we assume that the active and reactive power outputs of the storage device \(s\), denoted by \(P_s\) and \(Q_s\), respectively, are constrained by a maximum apparent power rating \(S_{s, \max}\). This constraint defines a \replaced{capability chart}{FOR} expressed as $\sqrt{P_s^2 + Q_s^2} \leq S_{s, \max}$.
This forms a convex nonlinear constraint, which may increase the computation time of optimization-based methods. To reduce computational complexity, we approximate the \replaced{capability chart}{FOR} using a regular polygon, as described in \cite{contreras2018} and as illustrated in Fig.~\ref{fig:devices_bess}. The active and reactive power variables are constrained to lie inside the green region of Fig.~\ref{fig:devices_bess}, mathematically referred to as \(\text{\replaced{CC}{FOR}}_{s}\).

Additionally, \acp{BESS} necessitate the management of their \ac{SOC} over time. Here, we model \ac{BESS} energy constraints with a lossless battery model:
\begin{equation}
\begin{aligned}
SoC_{s}^{0} &= SoC_{s}^{\text{init}}, \\
SoC_{s}^{t+1} &= SoC_{s}^{t} - \frac{\Delta t P_{s}^{t}}{C_{s}}, 
\end{aligned}
\end{equation}
where $SoC_{s}^{\text{init}}$ and $SoC_{s}^{t}$ are the state of charge of storage device $s$ at time step $0$ and $t$, respectively, $P_s^t$ is the active power delivered to the grid, $C_s$ denotes the energy capacity of the battery, and $\Delta t$ is the duration of a time step. The limited \ac{BESS} energy capacity constrains the \ac{SOC} to remain between $SoC_{s}^{min}$ and $SoC_{s}^{max}$, which may further constrain the \ac{BESS} flexibility region, \(\text{\replaced{CC}{FOR}}_{s}^{t}\).

\subsubsection{PV Systems}
\label{subsec:device_pv}
In contrast to battery storage systems, PV systems do not necessitate the modeling of intertemporal coupling, since they have no energy-related states such as stored energy or \ac{SOC} that evolve over time. 
Only technical power constraints restrict the power outputs $P_{pv}$ and $Q_{pv}$ of PV systems. Their maximum active power output varies throughout the day, due to time-varying solar irradiance.
The active power of the PV systems can be fully modulated between zero and their maximum active power. For reactive power, we assume controllability within a defined range of power factors, analogous to the approach in \cite{contreras2018}. These constraints are illustrated in the PQ-plane in Fig.~\ref{fig:devices_pv}, and are referred to as \(\text{\replaced{CC}{FOR}}_{pv}^{t}\).

\subsubsection{Controllable Loads}
\label{subsec:device_cl}
The third source of flexibility that we consider in this study is controllable loads, primarily heat pumps. In this paper, we define controllable loads as flexible loads characterized by a fixed power factor \cite{contreras2018}. Hence, the \replaced{capability chart}{FOR} of such devices, designated as \(\text{\replaced{CC}{FOR}}_{c}\) for a load $c$, is restricted to a line whose slope depends on the power factor of the device, as shown in Fig.~\ref{fig:devices_cl}.

Depending on the type of load, additional constraints apply. For instance, heat pumps' energy constraints also limit their flexibility. 
Indeed, heat pumps may consume more active power and store excess energy thermally, resulting in a higher indoor temperature. However, increased or decreased heat pumps' consumption must not lead to unacceptable temperatures for inhabitants, i.e., temperatures must stay in a predefined acceptable temperature range. In this paper, we adopt a simple intertemporal thermal model, where the room temperature $T_{c}^{t}$ at time $t$ controlled by heat pump $c$ is given by:
\begin{equation}
    \begin{aligned}
        T_{c}^{t+1} = T_{c}^{t} + \frac{P_{c}^{\text{flex},t}}{P_{c}^{\text{base},t}} q_{\text{heat}}^t \Delta t,
    \end{aligned}
    \label{equ:heating_model}
\end{equation}
where $q_{heat}^t$ designates the nominal heating demand, in °C/h, i.e., the thermal power required to keep the indoor temperature at the preferred temperature. The heating demand $q_{heat}^t$ is informed by \cite{lastprofil} and depends on the outside temperatures. It translates into a base active power consumption $P_{c}^{\text{base},t}$. For a deviation in active power consumption, $P_{c}^{\text{flex},t}$, we assume that the indoor temperature evolves proportionally to the ratio between the flexible and the baseline active power consumption. In case there is no deviation from baseline, i.e., $P_{c}^{\text{flex},t} = 0$, the temperature will remain unchanged.

In addition to heat pumps, we include electric boilers as controllable loads. However, since these systems are expected to be replaced by heat pumps in the future in Switzerland, they will play a smaller role in future distribution grid flexibility \cite{bfe2020energyperspectives}. Hence, we omit a detailed discussion of their modeling here and refer the reader to \cite{boiler_modeling} for the model of flexible deferrable appliances that is employed in the analysis.

\subsection{Distribution Grid Modeling}
The aggregated flexibility of a collection of flexible devices located in a distribution grid is restricted not only by their individual constraints but also by the technical constraints of the grid. 

Therefore, we must model the power grid and its constraints. \added{Since we subsequently integrate the power flow model into an optimization framework, we opt for a linear power flow approximation as opposed to a full non-linear AC model. This approach circumvents the computational complexity and optimality challenges associated with the nonlinearity and non-convexity of the AC power flow equations while it maintains sufficient accuracy for distribution networks \cite{bolognani2016}.} \deleted{In this paper,} We approximate the power flow equations using a standard first-order Taylor expansion of the nonlinear equations around a standard operating point characterized by a unity voltage magnitude and a zero angle. This linearly relates how a relative change in voltage angles $\theta$ and voltage magnitudes $\Delta U$ results in a change in active and reactive power injections, i.e.:
\begin{equation}
    \begin{bmatrix} P^{t} \\ Q^{t} \end{bmatrix} = \begin{bmatrix} P_{ref} \\ Q_{ref} \end{bmatrix} + 
\begin{bmatrix} J_{P\theta} & J_{PU} \\ J_{Q\theta} & J_{QU} \end{bmatrix} \begin{bmatrix} \theta^{t} \\ \Delta U^{t} \end{bmatrix}
\label{equ:powerflow}
\end{equation}
where the Jacobian submatrices are defined component-wise as:
\begin{equation}
    \begin{aligned}
    J_{P\theta}(i,k) &= \begin{cases} \sum_{j \ne i} -b_{ij} & \text{if } i=k \\ b_{ik} & \text{if } i \ne k \end{cases} \\
    J_{QU}(i,k) &= \begin{cases} - \sum_{j \ne i} 2b_{ij}^{sh} -b_{ij} & \text{if } i=k \\ b_{ik} & \text{if } i \ne k \end{cases}
     \\
    J_{PU}(i,k) &= -J_{Q\theta}(i,k) = \begin{cases} \sum_{j \ne i} g_{ij} & \text{if } i=k \\ -g_{ik} & \text{if } i \ne k \end{cases}, 
    \end{aligned}
    \label{equ:Jacobian}
\end{equation}
where $g_{ik}$, $b_{ik}$, and $b_{ik}^{sh}$ are the series conductance, series susceptance, and shunt susceptance, respectively, of the line from node $i$ to $k$. Here, we neglect shunt conductances for simplicity. Hence, the operating point power injections $P_{ref,i}$ and $Q_{ref,i}$ are $Q_{ref,i} = - \sum_{j \ne i} b_{ij}^{sh}$ and $P_{ref,i} = 0$. Furthermore, we define the PCC as the slack bus constraint, with fixed voltage magnitude of 1~p.u. and angle of 0 degrees. \added{The accuracy of the linear model was validated against a nonlinear AC power flow solution over a 24-hour historical simulation of the case study. The results yielded a mean absolute error of 0.0001~p.u. for nodal voltage magnitudes and $0.004^\circ$ for voltage angles. While approximation errors typically increase under more stressed operating conditions, such linearizations remain sufficiently accurate for the purpose of flexibility quantification in this study \cite{contreras2018, bolognani2016, nerowski2024}.}

Two types of grid constraints may restrict distribution grid flexibility. First, line limits restrict the maximum apparent power flowing in the grid, formally defined as ${\sqrt{{P_{ik}^{t}}^2 + {Q_{ik}^{t}}^2} \leq S_{ik,max}}$. Similarly to the constraints on \acp{BESS}, we reformulate these nonlinear constraints, using regular polygons \cite{contreras2018}. Second, net power injections at all nodes are restricted by voltage limits, delimited by $U_\text{min}$ and $U_\text{max}$. 
\begin{figure}
    \centering
    \includegraphics[width=0.6\linewidth]{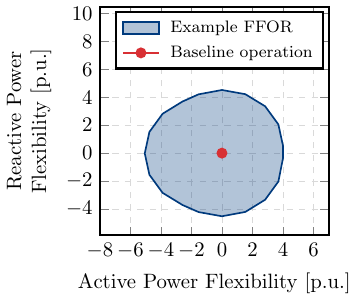}
    \caption{Example of an FFOR at the PCC.}
    \label{fig:examples_for_ffor}
\end{figure}
 \subsection{Feasible Flexibility Operating Region (FFOR)}

\subsubsection{Definition}
The FOR designates the feasible active and reactive power exchange values at the PCC \cite{riaz2019}, \cite{contreras2019comparison}, \cite{lopez2021quickflex}. However, in the flexibility context, we may be more interested in the deviation from a planned operating point than in an absolute power exchange value.  
Consequently, we introduce the concept of the \ac{FFOR}, which represents all possible active and reactive power deviations from the baseline operation. By definition, the FFOR is centered around the origin. The origin of the FFOR corresponds to no flexibility provision, i.e., the operating point does not deviate from the planned schedule. Fig.~\ref{fig:examples_for_ffor} shows an example of an FFOR. It differs from the FOR only in the location in the PQ-plane, but has the same shape. \added{Consequently, the FFOR is also not theoretically guaranteed to be convex. For instance, in \cite{silva2018}, the FOR is shown to be potentially disjoint in the presence of strong discontinuous variables, while the study in \cite{contreras2019} shows that a linearized power flow can lead to an FOR containing operating points for which the corresponding AC power flow is infeasible. However, most empirical results demonstrate that for typical network layouts and a larger number of distributed devices without individual strong nonconvexities, the resulting FOR exhibits near-convex behavior \cite{riaz2019, contreras2019}. For the purpose of this study, we therefore assume the FFOR to be practically convex and model its computation via a \ac{LP}, which ensures convexity within the optimization framework.}

\subsubsection{One-Timestep FFOR}
To compute the FFOR, we apply an iterative OPF-based algorithm \cite{lopez2021quickflex}. At timestep $t$, we approximate the set of feasible power exchanges at the PCC with a polytope of an increasing number of vertices. In brief, we iteratively solve an optimization that identifies, along a direction, i.e., a pre-defined PQ-line, the largest flexible power. For each direction, the optimal solution defines a vertex. The set of resulting vertices forms a polytope that approximates the FFOR. The optimization problem for timestep $t$ and a specific direction is given by
\begin{subequations}
\begin{align}
    \min_{P_{\text{pcc}}^t, Q_{\text{pcc}}^t} \quad & \alpha \left( P_{\text{pcc}}^t - P^{\text{base},t}_{\text{pcc}} \right) + \beta \left( Q_{\text{pcc}}^t - Q^{\text{base},t}_{\text{pcc}} \right) \\
    \text{s.t.} \quad 
    & (P_s^{t,k}, Q_s^{t,k}) \in \text{\replaced{CC}{FOR}}_s^{t,k}, \hspace{1cm} \forall s \ \forall k \label{subeq:oneTimeOpt-s}\\
    & (P_{pv}^{t,k}, Q_{pv}^{t,k}) \in \text{\replaced{CC}{FOR}}_{pv}^{t,k}, \hspace{1cm} \forall pv \ \forall k \label{subeq:oneTimeOpt-pv}\\
    & (P_c^{t,k}, Q_c^{t,k}) \in \text{\replaced{CC}{FOR}}_c^{t,k}, \hspace{1cm} \forall c \ \forall k\label{subeq:oneTimeOpt-c}\\
    & \text{Power flow model,} \\
    & \text{Line constraints,}
    \label{subeq:oneTimeOpt-l}
\end{align}\label{eq:oneTimestepOpt}
\end{subequations} \par\noindent 
where $k$ denotes the different buses in the network. At each iteration, the weighting factors $\left(\alpha, \beta\right)$ define a specific direction. For instance, $\left(1, 0\right)$ defines a direction along the $P$-axis which aims to maximize the active power. We iteratively explore directions until we can no longer observe any significant increase in the polytope area.

\subsubsection{Multi-Timesteps FFOR} 
A key advantage of the FFOR over the FOR is its ability to visually represent flexibility across multiple timesteps. In the FOR, the visual identification of a constant power deviation is complex, as the operating point changes throughout the horizon and the region shifts in the PQ plane. In contrast, a constant power deviation in the FFOR is described by a fixed point. Hence, the FFOR facilitates the identification of metrics spanning across multiple timesteps and time-varying baselines. 

Over multiple timesteps, additional time-coupling energy constraints must be added to (\ref{eq:oneTimestepOpt}). Given a sustained duration $d$, we can determine the multi-timesteps FFOR as the constant active and reactive power deviation values that can be sustained over $d$. Using the same iterative procedure as in (\ref{eq:oneTimestepOpt}), we solve the following optimization:
\begin{subequations}
\begin{align}
    \min_{P_{\text{pcc}}, Q_{\text{pcc}}} \quad & \alpha P_{\text{pcc}}^{\text{flex},d} + \beta Q_{\text{pcc}}^{\text{flex},d} \\
    \text{s.t.} \quad 
    & \text{SOC \& Temperature Constraints}, \\
    & (\ref{subeq:oneTimeOpt-s})-(\ref{subeq:oneTimeOpt-l}), \hspace{2cm} \forall t=0, \cdots, d, \\
    & P_{\text{pcc}}^{\text{flex},t} = P_{\text{pcc}}^t - P^{\text{base},t}_{\text{pcc}}, \quad \forall t=0, \cdots, d, \label{subeq:multiTimeP} \\
    & Q_{\text{pcc}}^{\text{flex},t} = Q_{\text{pcc}}^t - Q^{\text{base},t}_{\text{pcc}}, \quad \forall t=0, \cdots, d. \label{subeq:multiTimeOptQ} \\
    & P_{\text{pcc}}^{\text{flex},t} = P_{\text{pcc}}^{\text{flex},t+1}, \hspace{0.8cm}  \quad \forall t=0, \cdots, d-1,\label{subeq:multiTimeOptSustainedP}\\
    & Q_{\text{pcc}}^{\text{flex},t} = Q_{\text{pcc}}^{\text{flex},t+1}, \hspace{0.8cm}  \quad \forall t=0, \cdots, d-1, \label{subeq:multiTimeOptSustainedQ}
\end{align}
\label{eq:multiTimestepOpt}
\end{subequations}
where \eqref{subeq:multiTimeOptSustainedP} and \eqref{subeq:multiTimeOptSustainedQ} ensure the sustained duration of a constant flexibility at the PCC over multiple timesteps. \added{Note that this optimization framework can be readily extended to other types of flexibility sources characterized by different convex capability regions and time-dependent constraints. However, certain resources are anticipated to introduce unique modeling challenges. For instance, electric vehicles exhibit discontinuous behavior due to connecting and disconnecting events \cite{taheri2022}, and other distributed assets may necessitate the explicit modeling of operational uncertainties in their flexibility quantification \cite{capitanescu2021}.}

\section{Case Study: Walenstadt Distribution Grid}\label{sec:caseStudy}
\added{The utility \textit{Wasser- und Elektrizitätswerk Walenstadt} (WEW) supplies electricity to more than 5000 people in Walenstadt. It operates the local distribution grid that hosts a large number of PV installations, heat pumps, and multiple utility-scale \acp{BESS}. The grid also includes three small run-of-the-river power plants, which are assumed to be nonflexible power production units in this work. \added{As part of the \textit{Grid2050} initiative, WEW and a growing community of its customers collaborate with Swiss research institutes to advance future distribution networks. \cite{grid2050} This partnership is currently developing an in-vivo testbed featuring controllable household assets, and it provides the real-world data and network topology upon which this case study is based.}}
In this section, we introduce the local distribution grid of Walenstadt in detail. We specifically explain how we leverage the historical measurement data to model a realistic case study.

\subsection{Topology and Grid Information}
\begin{figure}
    \centering
    \includegraphics[width=\linewidth]{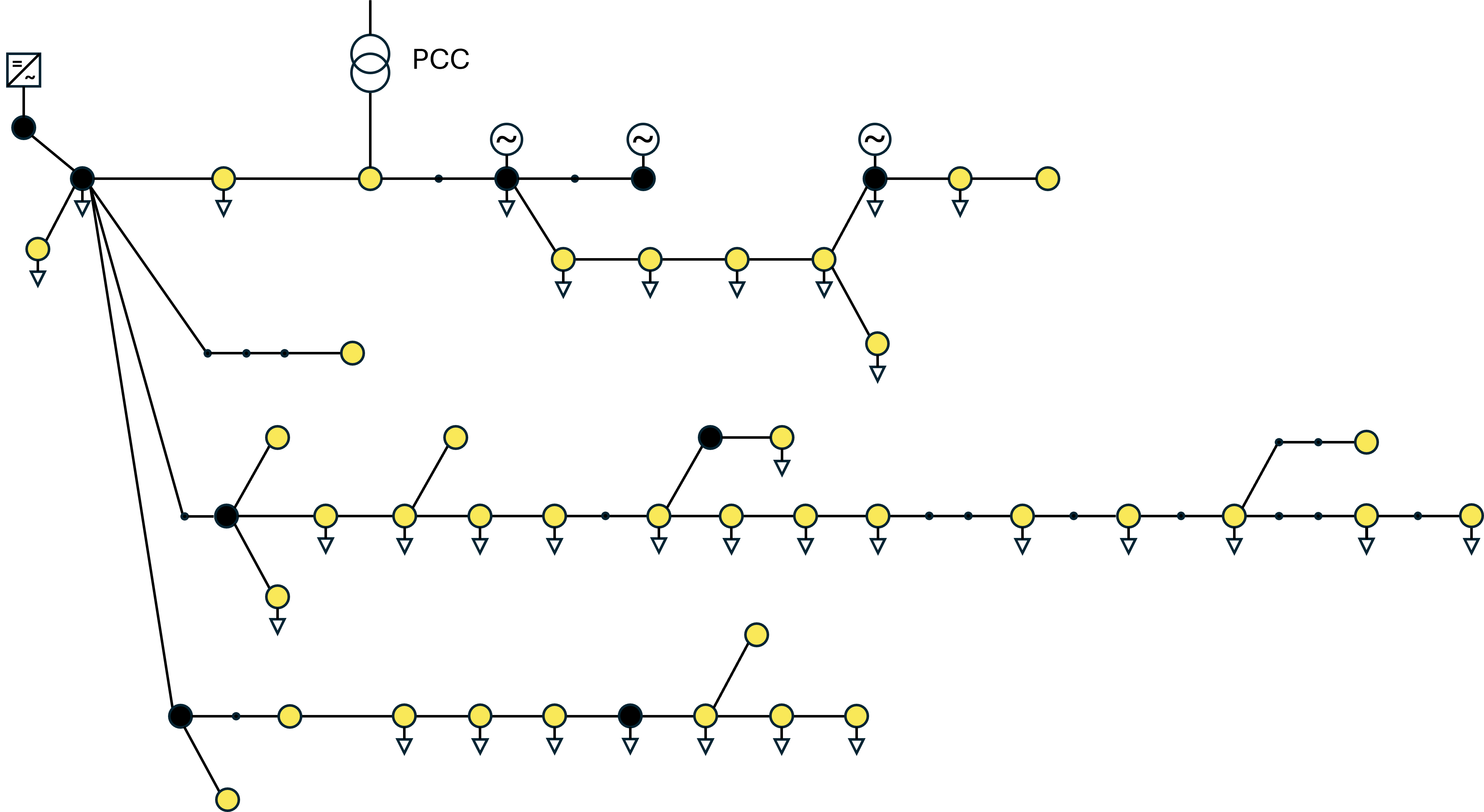}
    \vspace{1.5pt}
    \caption{Walenstadt grid: The nodes represent either connection points for the large \ac{BESS} (top left) or run-of-river power plants, or represent transformer substations to the \ac{LV} grid. Controllable loads are represented by triangles, and yellow-colored nodes indicate installed PV capacity in the downstream network.}
    \label{fig:topology}
\end{figure}

The local distribution network consists of an \ac{MV} level with a nominal voltage of 16~kV and a \ac{LV} level at 400~V. Fig.~\ref{fig:topology} schematically depicts the \ac{MV} grid, which exhibits a typical radial structure. This layout includes a single \ac{PCC} with the \ac{HV} grid, at which we evaluate the aggregated flexibility of the entire distribution grid.
\begin{table}
\centering
\caption{Overview of Data Measurements and Resolutions}
\label{tab:data_measurements}
\begin{tabular}{|l|c|c|}
\hline
\textbf{Measurement} & \textbf{Time resolution} & \textbf{Spatial resolution} \\ \hline
\multicolumn{3}{|l|}{\textbf{Transformer substations}} \\ \hline
\quad Net active power & 15 min & Node \\
\quad Net apparent power & 15 min & Node \\
\quad Voltage magnitudes & 15 min & Node \\
\quad Current magnitudes & 15 min & Node \\ \hline
\textbf{PV generation (Total)} & 15 min & Grid \\ \hline
\textbf{MV hydropower} & 15 min & Node \\ \hline
\multicolumn{3}{|l|}{\textbf{Weather data}} \\ \hline
\quad Ambient temperature & $\leq$ 15 min & Grid \\
\quad Irradiance & $\leq$ 15 min & Grid \\ \hline
\end{tabular}
\end{table}

Using data provided by the local utility, WEW, we specify all line parameters indicated in (\ref{equ:Jacobian}), as well as the local line ratings that limit power flows in the grid. Furthermore, we assume uniform voltage limits of $U_\text{min} = 0.95$~p.u. and $U_\text{max} = 1.05$~p.u in the network. In contrast to the \ac{MV} grid, data on the \ac{LV} grid are not available. Therefore, the \ac{LV} level is modelled as an aggregated load profile at each transformer, assumed to represent the behavior of the underlying \ac{LV} grid. \added{Consequently, the specific impacts of the \ac{LV} network are not assessed. This limitation will be addressed in future work using newly available data from the \textit{Grid2050} project.}

\subsection{Device-Level Information}
As presented in Section~\ref{sec:methodology}, our methodology requires the identification of the technical characteristics of flexible devices, as well as baseline operation, in order to quantify flexibility. In Walenstadt, three large-scale batteries are installed, and we assume that one of these batteries, with a capacity rating of 4~MW/5~MWh, is available to support distribution grid flexibility in a SOC range from 0.4 to 0.8.\footnote{We assume that this battery is the only \ac{BESS} participating in the distribution grid flexibility in Walenstadt.} Additionally, there are many distributed small-scale \ac{PV} systems and heat pumps. Nevertheless, to date, their exact technical properties and locations in the grid are unknown to the utility.

Hence, alongside a real-life grid topology, we must identify realistic device-level information. For this purpose, we rely on historical measurement data provided by the utility of Walenstadt, for which an overview is presented in Table~\ref{tab:data_measurements}. Based on these data, we identify flexible devices' individual technical properties, specifically PV systems' and heat pumps' power capacity.

\begin{figure}
    \centering
    \includegraphics[width=0.9\linewidth]{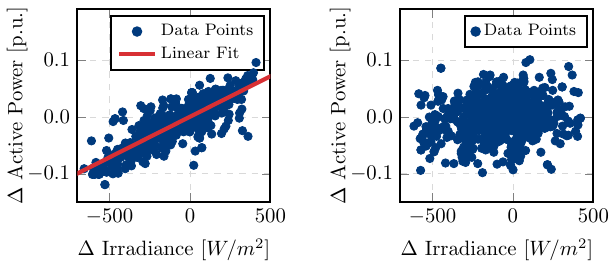}
    \caption{Illustration of the linear relationship between irradiance and net active power at different nodes. Left: Data and linear fit for node \textit{A}, showing a significant linear relationship, indicating a large amount of \ac{PV}. Right: Data for node \textit{B}, showing no significant linear relationship, indicating no \ac{PV}.}
    \label{fig:PVregression}
\end{figure}

\begin{figure}
    \centering
    \includegraphics[width=0.9\linewidth]{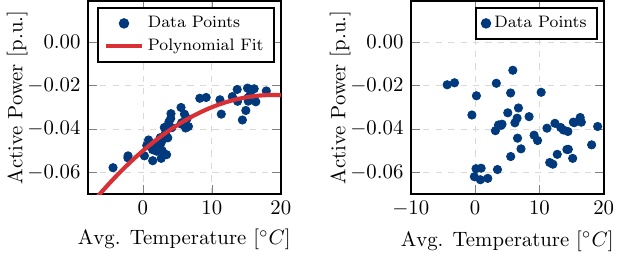}
    \caption{Illustration of the polynomial relationship between temperature and net active power. Left: Data and polynomial fit for node \textit{A}, showing a significant regression fit, indicating a moderate amount of heat pump capacity. Right: Data for node \textit{C}, showing no significant polynomial relationship, indicating no to very little heat pump capacity.}
    \label{fig:HPregression}
\end{figure}

Based on the aggregated active power consumption measured at each node, we want to identify how many PV systems and heat pumps are connected to the node. To this aim, we rely on correlations. For PV systems, we assess the correlation between the net active power of a node and solar irradiance. As illustrated in Fig.~\ref{fig:PVregression}, a strong positive correlation indicates the presence of a high PV capacity. We know the total PV capacity installed in the grid, and we distribute it to the nodes proportionally to the identified correlation. For heat pumps, we analyze the correlation between the node-wise average net active power and the average ambient temperature during periods with no solar radiation, as depicted in Fig.~\ref{fig:HPregression}. We specifically fit a second-order polynomial to infer the node-wise installed heat pump capacity. The node-wise heat pump capacity can be estimated as the difference in net power at no heating and full heating power. For an appropriate heat pump sizing, a heat pump should heat at full power at an ambient temperature of -8°C and should not consume any power at 15°C \cite{heizkurve}.

\subsection{Baseline Operation}
As the FFOR describes feasible deviations from planned operation, we must provide an estimate of the grid baseline operation in Walenstadt. Based on the estimated node-wise capacity, we derive the baseline operation of the PV systems and heat pumps, using a few assumptions. First, we assume that PV systems track the maximum power generation at a unity power factor, in baseline operation. Second, we assume that the baseline active power consumption of the heat pumps aims to keep room temperatures constant, which corresponds to the load profiles adapted from the heating demand described in \cite{lastprofil}. At each node, the nonflexible load is deduced as the remaining load after subtracting the heat pumps' consumption and PV systems' production. Finally, we assume that the large \ac{BESS}' operation entirely supports distribution grid flexibility and thus has a zero consumption/generation baseline.

\section{Results}\label{sec:results}
In this section, we provide a comprehensive analysis of the flexibility potential of the distribution grid operated by the utility of Walenstadt. In the remainder, the terms \textit{positive flexibility} and \textit{negative flexibility} refer to an increase in generation or a decrease in consumption of active power, and to a decrease in generation or an increase in consumption of active power, respectively.

\subsection{Illustrative Example}
\begin{figure}
    \centering
    \includegraphics[width=\linewidth]{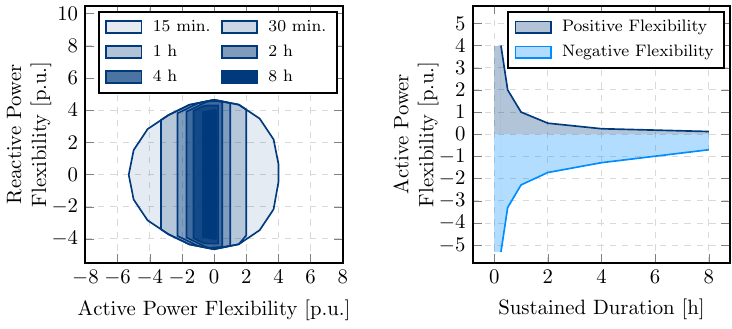}
    \caption{Quantification of flexibility on 1 September 2021 at 14:00. Left: The FFOR for different durations of sustained flexibility provision. Right: Maximum active power flexibility over different durations of sustained provision.}
    \label{fig:FFOR_base_steps}
\end{figure}

First, we illustrate the flexibility potential of WEW's distribution grid on a specific date, namely on 1 September 2021 at 14:00. 
Fig.~\ref{fig:FFOR_base_steps} depicts the single-timestep FFOR (15-minute) and the multi-timestep FFOR for different sustained durations. We observe that the \ac{FFOR} is not symmetric with respect to active power flexibility. This can be attributed to three main reasons. Firstly, the active power flexibility of different types of devices is asymmetric. For instance, a PV system can only curtail, but not increase generation, since its baseline operation already corresponds to a maximum power operation. Secondly, the operation of some devices is time-dependent, e.g., because they are influenced by the weather. A third reason may be operational constraints in the grid, which can impose asymmetric limits on the \ac{FFOR}.

Fig.~\ref{fig:FFOR_base_steps} also reveals that the active power flexibility, that can be sustained over a certain duration, reduces nonlinearly with that duration. This is due to some of the flexible devices being energy-constrained. For instance, the positive flexibility shown in Fig.~\ref{fig:FFOR_base_steps} is mainly provided by the large \ac{BESS}. Its available amount of energy, i.e., the amount of kWh that can be charged or discharged without reaching the battery's \ac{SOC} limits, is constrained. Since this amount needs to be dispersed over the whole horizon of provision, the sustained power level decreases. Generally, short-term flexibility is constrained by capacity constraints, as indicated by the round shape of the FFOR for the duration of a quarter hour, while energy-related constraints increasingly characterize the FFOR over longer durations.
\subsection{Flexibility Gain from Heat Pump Integration}
\begin{figure}
    \centering
    \includegraphics[width=1\linewidth]{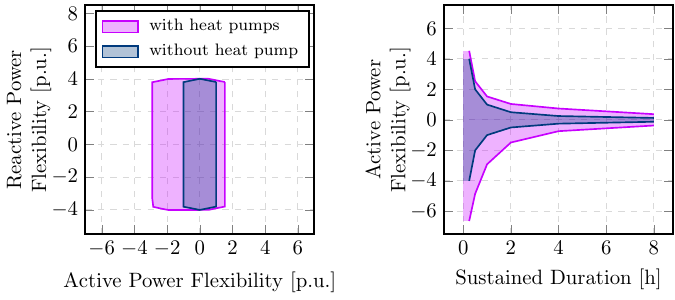}
    \caption{Quantification of flexibility on 1 November 2021 at 19:00, illustrating the flexibility gain of heat pumps. Left: The FFORs for a sustained duration of 1 hour. Right: Active power flexibilities over different sustained durations.}
    \label{fig:flexibility_gain}
\end{figure}
To assess the impact of integrating small-scale distributed flexibility-providing assets, we consider two scenarios: one with the consideration of heat pumps as flexibility providers, and one without. Fig.~\ref{fig:flexibility_gain} illustrates the multi-timestep FFOR, sustained over 1~hour, and the maximum active power that can be provided in both directions for different sustained durations, for a scenario on a November evening in 2021. It is evident that the flexibility potential from small-scale assets like heat pumps is significant. Specifically, heat pumps increase the positive and negative active power flexibility by around 0.5~p.u. and 1~p.u., respectively, over the duration of two hours. This amount even exceeds the capacity of the \ac{BESS} that was included in the study.

Interestingly, the flexibility gain for short-term flexibility of up to 3 hours of sustained duration is asymmetric. Reductions in consumption are limited by baseline demand, and increases in consumption are limited by heat pump capacity or grid constraints. For longer periods, energy constraints come into play, and sustained operation at maximum or minimum consumption is not possible anymore, since this would overheat or cool the rooms too much. Thus, flexibility provision is closely tied to baseline consumption, which in turn strongly depends on time of day and season.

\subsection{Seasonal Availability of Flexibility}
\begin{figure}
    \centering
    \includegraphics[width=1\linewidth]{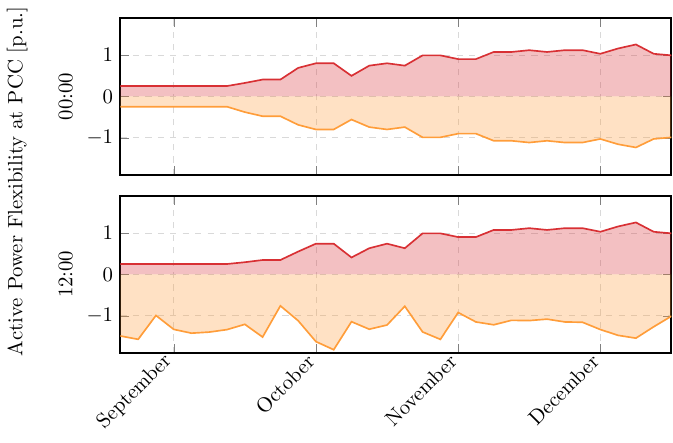}
    \caption{Seasonal evaluation of active power flexibility for a sustained duration of 4 hours.}
    \label{fig:seasonal}
\end{figure}

Distribution grid flexibility exhibits significant temporal variability. To assess this variability and to provide insights into periods of high and low flexibility potential, we quantify the maximum active power flexibility for a sustained 4-hour duration for a range of different days. Fig.~\ref{fig:seasonal} presents the results of this analysis, conducted from September to December, starting the quantification at 12~a.m. and 12~p.m. every day.

We observe that the amount of active power flexibility varies significantly with the season. Flexibility increases during the colder months because flexibility from heat pumps is only available when they are in operation. Positive flexibility from reduced consumption can only be provided if the baseline consumption is significant, i.e., when heating is required. Conversely, negative flexibility can only be provided in cold weather conditions, where increased consumption does not cause rooms to overheat too quickly.
In contrast, the flexibility potential is lowest during the warmer months, especially at night. This is because there are no large loads, such as heat pumps, in the network that can be adjusted to provide flexibility. Consequently, nighttime flexibility in September relies entirely on the \ac{BESS}.
In addition to these seasonal variations, there are also daily changes in flexibility. In conclusion, the analysis shows that while there exists a significant flexibility potential, this potential is time varying. At times, there is even a lack of flexibility, which can have implications for participation in frequency regulation markets or optimal planning of grid operations or reinforcement.

\subsection{Limitations to Flexibility in Future Scenarios}
\begin{figure}
    \centering
    \includegraphics[width=1\linewidth]{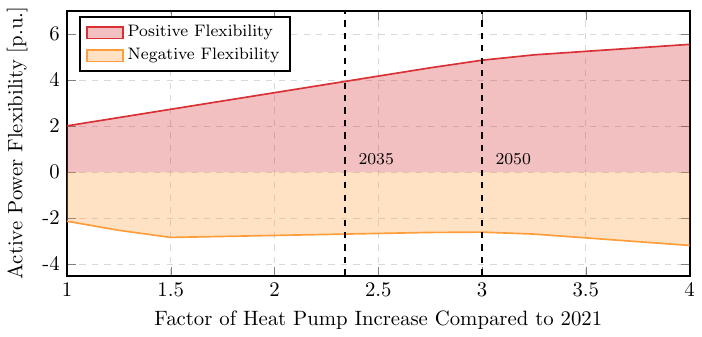}
    \caption{Evolution of active power flexibility sustained for 2 hours with increasing number of installed heat pumps. The numbers for other types of devices are based on projected values for 2035 from the Swiss Federal Office of Energy \cite{bfe2020energyperspectives}. The simulated scenario takes place on a projected typical day in December 2035 at 20:00.}
    \label{fig:future}
\end{figure}

While the current distribution grid flexibility is of interest, assessing the evolution of future potential and its limitations is just as important. Generally, the number of devices capable of providing flexibility is expected to increase significantly. Fig.~\ref{fig:future} shows the available active power flexibility for a duration of 2 hours for increased levels of heat pump penetration. It indicates that this growth in the number of flexible devices generally translates into increased flexibility at the PCC. The rise in heat pump installations leads to an increased baseline consumption during winter, allowing for greater consumption reductions when needed, i.e., an increased positive flexibility, and a higher installed power capacity, i.e., an increased negative flexibility.

While flexibility at the PCC generally increases with more flexible devices, it does not necessarily scale linearly. In Fig.~\ref{fig:future}, the positive flexibility shows a steady, monotonic increase. Initially, it grows linearly with heat pump capacity, as the baseline consumption scales linearly. This increase continues until operational constraints, such as line capacity limits, are reached under baseline operation. When a line becomes constrained under baseline operation, the installation of a further heat pump in an area downstream can no longer increase the baseline power drawn from the PCC. Since this effect does not occur on all lines or feeders simultaneously, positive flexibility can still increase, though at a reduced rate.
A similar reasoning applies to negative flexibility. It does not always increase monotonically and can even decrease. In the example shown in Fig.~\ref{fig:future}, this is due to limited line flow capacity. As baseline consumption approaches line limits because of the increasing number of heat pumps, the ability to consume additional power as negative flexibility decreases.

In conclusion, we can state that an increasing number of flexible devices does not necessarily translate to a proportional increase in flexibility at the PCC. Instead, the distribution of devices in the grid and the grid topology play a crucial role. Essentially, the flexibility at the PCC is topology-dependent. Even a single constraining line, especially if located upstream in a feeder, can prevent the translation of device-level flexibility to grid-wide flexibility at the PCC.

\section{Conclusion}\label{sec:conclusion}
In this paper, we study the aggregated flexibility potential of distribution grids on the real grid of Walenstadt, a municipality in Eastern Switzerland. We demonstrate that incorporating devices such as heat pumps and PV systems significantly expands the FFOR. Aggregating these devices can yield a flexibility gain comparable to that of large-scale battery energy storage systems. In view of the current impetus to integrate more PVs at the household level and to replace old heating systems with heat pumps, this finding has profound implications for the Swiss energy system. As more and more such devices find their way into the grid, the flexibility potential of distribution grids is expected to grow considerably. However, we also show that distribution grid flexibility is not symmetric or constant, but time-varying and dependent on the season. In fact, there are times with considerably less flexibility potential. Furthermore, simulations of future scenarios reveal that aggregated flexibility does not increase linearly or monotonically with higher levels of heat pump penetration. This nonlinearity is primarily due to the overloading of individual feeders, underscoring the significant impact of grid topology and network constraints on the overall aggregated flexibility potential. \added{These findings, which showcase the potential of small-scale flexibility and demonstrate the limitations imposed by temporal variability and grid constraints, provide vital real-world insights for both distribution and transmission grid operators to inform the design of flexibility strategies.} 

\added{While this study quantifies a technical upper-bound for aggregated flexibility, it does not investigate the impact of individual customer behavior or demand response mechanisms. Furthermore, integrating these results into real-time operation requires treating flexible assets within a closed-loop environment. This necessitates the continuous re-evaluation of aggregated flexibility, which will be further explored in upcoming projects.} Future research will also examine the trade-offs between using device-level flexibility to defer grid reinforcement and pursuing strategic reinforcement to increase distribution grid flexibility.

\section*{Acknowledgments}
This work was supported by the NCCR Automation, a National Centre of Competence in Research, funded by the Swiss National Science Foundation (grant number 51NF40\_225155), and was conducted in collaboration with the utility \textit{Wasser- und Elektrizitätswerk Walenstadt} under the \textit{Grid 2050} initiative.

\end{document}